\documentstyle[12pt,aaspp4]{article}

\def\be{\begin{equation}}
\def\ee{\end{equation}}
\def\ba{\begin{eqnarray}}
\def\ea{\end{eqnarray}}

\journalid{VOL}{JOURNAL DATE}
\articleid{START PAGE}{END PAGE}
\paperid{MANUSCRIPT ID}

\slugcomment{To appear in the Astrophysical Journal}

\lefthead{Friedman, Morsink}
\righthead{Axial instability of rotating relativistic stars}

\begin{document}

\title{Axial instability of rotating relativistic stars}
\author{John L. Friedman and Sharon M. Morsink}
\affil{University of Wisconsin-Milwaukee, P.O. Box 413, Milwaukee, WI
53201 
 \\ friedman@thales.phys.uwm.edu, morsink@pauli.phys.uwm.edu}
 
\begin{abstract} 
Perturbations of rotating relativistic stars can be classified by their
behavior under parity.  For axial perturbations (r-modes), initial data
with negative canonical energy is found with angular dependence
$e^{im\phi}$ for all values of $m\geq 2$ and for arbitrarily slow
rotation.  This implies instability (or marginal stability) of such
perturbations for rotating perfect fluids.  This low $m$-instability is
strikingly different from the instability to polar perturbations, which
sets in first for large values of $m$.  The timescale for the axial
instability appears, for small angular velocity $\Omega$, to be
proportional to a high power of $\Omega$.  As in the case of polar
modes, viscosity will again presumably enforce stability except for
hot, rapidly rotating neutron stars.

This work complements Andersson's numerical investigation of axial modes in slowly rotating stars.
\end{abstract}


\section{Introduction}
\label{s:introduction}

A rotating star is invariant under reflection in the equatorial plane
and rotations about the axis of symmetry.  It is therefore invariant
under parity, the composition of reflection in the equatorial plane and
rotation by $\pi$, and perturbations of a rotating star can be chosen
to have definite parity.  That is, the space of solutions to the
Einstein-perfect-fluid equations, linearized about the equilibrium
star, is a direct sum of dynamically preserved subspaces with parity
eigenvalues $\pm 1$.  \footnote{ It would be natural to call these
``odd'' or ``even'' perturbations, but some of the literature in
relativistic astrophysics adopts the Regge-Wheeler (1957) terminology, using
the term ``odd-parity'' not for a perturbation that changes sign under
parity but for an axial perturbation, whose behavior is opposite to
that of $Y_l^m$.}

Perturbations of a spherical star can be divided into two classes,
axial and polar, depending on their behavior under parity.  Where polar
tensor fields on a 2-sphere can be constructed from the scalars
$Y_l^m$, their gradients $\nabla_A Y_l^m$, and the metric $e_{AB}$
on the 2-sphere, axial fields involve the pseudo-tensor
$\epsilon_{AB}$, and their behavior under parity is opposite to that of
$Y_l^m$.  Thus axial perturbations of odd $l$ are invariant under
parity, and axial perturbations with even $l$ change sign.  If a mode
varies continuously along a sequence of equilibrium configurations that
starts with a spherical star and continues along a path of increasing
rotation, the mode will be called axial if it is axial for the
spherical star.  Its parity cannot change along the sequence, but $l$
is well-defined only for modes of the spherical configuration.  

Axial perturbations of a spherical stellar model are time-independent
convective currents that do not change the density and pressure of the
star and do not couple to gravitational waves. 
Along a continuous sequence of relativistic perfect-fluid
models, a mode changes from stable to unstable when its frequency is
zero; because they have vanishing frequency in nonrotating models,
axial modes could thus be unstable for any model with nonzero
rotation.  Andersson's numerical work and the computation we present
here shows that this is the case:  Axial modes for every value of $m$
are unstable for arbitrarily slow rotation.

In spherical stars, gravitational radiation removes positive angular
momentum $L_z$ from a mode moving in the positive $\phi$ direction and
negative angular momentum from a backward-moving mode; and it
therefore damps all time-dependent nonaxisymmetric modes.  Once the
angular velocity of the star is sufficiently large, however, a mode
that moves backward relative to the star is dragged forward relative to
an inertial observer.  Gravitational radiation will then remove
positive angular momentum from the mode.  But a mode that moves
backwards relative to the fluid has negative angular momentum, because
the perturbed fluid does not rotate as fast it did without the
perturbation.  The radiation thus removes positive angular momentum
from a mode whose angular momentum is negative.  By making the angular
momentum of the perturbation increasingly negative, gravitational
radiation drives the mode (Friedman and Schutz 1978, Friedman 1978, 
Chandrasekhar 1970).  

For slowly rotating Newtonian stars, the frequencies of axial modes are
known (see Sect. \ref{s:newton}), and they lie in this unstable regime.
In general relativity, Andersson (1997) numerically computes the real
part of the frequency of these modes in a slow-rotation approximation,
while we show analytically that that their canonical energy can be made negative; the fact that they are unstable (or
marginally stable) follows analytically from this and numerically from
Andersson's frequency.

Formal work on the stability of rotating stars does not
distinguish between axial and polar modes, but explicit studies of
nonaxisymmetric instability of stellar modes had been restricted to
polar perturbations, for which the onset of instablity is very
different. There are unstable polar modes with arbitrarily small
rotation, but as $\Omega\rightarrow 0$, the minimum $m$ for which a
mode is unstable grows without bound.

In Sect. \ref{s:newton}, we consider the stability of axial
perturbations in the Newtonian limit,  showing that for any nonzero
$\Omega$ the canonical energy can be made negative for each $m\geq 2$.
The perturbations we consider have angular behavior $e^{im\phi}$ and
parity $(-1)^{m+1}$, corresponding to an $l=m$ mode of a corresponding
sperical star.  In Sect. \ref{s:gr}, we extend the computation to
general relativity, again finding initial data with negative canonical
energy. Finally, in Sect. \ref{s:time}, we estimate the timescale and
briefly discuss implications and future problems.
 
Spacetime indices will be lower-case Greek letters, spatial indices lower-case Latin.  We will work in gravitational units, setting $G=c=1$.  

\section{Newtonian stars}
\label{s:newton}

We consider a uniformly rotating, axisymmetric, self-gravitating 
perfect fluid, described by a gravitational potential $\nu$, density $\rho$,
pressure $p$ and 3-velocity 
\be
v^a = \Omega \phi^a.
\ee
These satisfy the equilibrium equations
\be
\nabla^2\nu=4\pi\rho
\ee

\be
\nabla_a (h-\frac12 v^2 + \nu )=0, 
\label{equil}\ee
together with an equation of state of the form
\be
\rho = \rho(p).
\ee
Here $h$ is the specific enthalpy in a comoving frame, 
\be
h = \int \frac{dp}\rho.
\ee

We will work in the Lagrangian formalism (Friedman and Schutz 1978a)\markcite{F&S75a} describing perturbations in 
terms of a Lagrangian displacement $\xi^a$ that connects fluid elements 
in the equilibrium and perturbed star.  The Eulerian change $\delta Q$ in a quantity $Q$ is related to its Lagrangian change $\Delta Q$ by 
\begin{equation}
\Delta Q =\delta Q + \mbox{\pounds}_\xi Q,
\end{equation}
where $\mbox{\pounds}_\xi$ is the Lie derivative along $\xi^a$. 

The displacement $\xi^a$ determines the fluid perturbation:
\begin{equation}
\Delta v^a = \partial_t\xi^a,
\end{equation}
\begin{equation}
\frac{\Delta p}{\gamma p} = \frac{\Delta\rho}{\rho} =-\nabla_a\xi^a .
\end{equation}
For angular dependence $e^{im\phi}$, we have the corresponding Eulerian
changes
\begin{eqnarray}
\delta v^a &=& (\partial_t+im\Omega )\xi^a\\
\delta\rho &=& -\nabla_a (\rho\xi^a)\\
\delta p &=& \frac{dp}{d\rho} \delta\rho ;
\end{eqnarray}
and the change in the potential is given by
\begin{equation}
\nabla^2\delta\nu =  4\pi\, \delta\rho .
\end{equation}

An axial perturbation of a spherical Newtonian star is a change in the
fluid's velocity of the form
\begin{equation}
\delta v^a= \tilde\zeta(r)\ \epsilon^{abc}\nabla_b r\nabla_c Y_l^m.
\label{deltav}\end{equation} 
That is, it is a vector with arbitrary $r$-dependence, whose behavior
on a symmetry two-sphere is given by $\epsilon^{AB}\nabla_B Y^l_m$.
\footnote{One can regard a tensor on the two-sphere as a tensor on
$R^3$ that is is orthogonal in all of its indices to $\nabla_a
r$ and is Lie derived by $\nabla^a r$.  The antisymmetric tensor
$\epsilon_{AB}$ associated with the unit metric $e_{AB}$ on the
two-sphere is then identified with $\epsilon_{ab} =
\epsilon_{abc}\nabla^a r$, and $\epsilon^{AB}\nabla_B Y_l^m$ is
identified with $\epsilon^{ab}\nabla_b Y_l^m$.}

The time independence of the perturbation can be seen as follows.
An axial displacement vector has the form
\begin{equation}
\xi^a =\zeta (r,t) \epsilon^{abc}\nabla_b r\nabla_c Y_l^m .
\label{xi}\end{equation}
The perturbed scalars ($\delta p$, $\delta \rho$,
$\delta\nu$) must vanish, because they are proportional to $Y_l^m$ in
an $(l,m)$ representation and thus cannot be axial.  Explicitly, from
the relations
\begin{equation}
\nabla_a\xi^a=0 , \hspace{3mm} \xi^a\nabla_a\rho =0 ,
\end{equation}
it follows that the Eulerian change in $\rho$---and hence in $p$ and $\nu$---vanishes.  
Then only the velocity changes, with
\begin{equation}
\delta v^a=\partial_t\xi^a ;
\end{equation}
and the perturbed Euler equation,
\begin{equation}
\delta [(\partial_t+\mbox{\pounds}_v)v_a
        +\nabla_a(h-\frac{1}{2} v^2 + \nu )] = 0 ,
\label{euler}\end{equation}
becomes simply
\begin{equation}
\partial_t\delta v^a=0 .
\end{equation}
Thus for a spherical Newtonian star, axial perturbations are
time-independent changes of the velocity, having the form (\ref{deltav}).

  We will shortly derive the axial instability in a way that does not
rely on normal modes.  Let us first however note that the known frequency
of axial modes suffices to imply their instability for slowly rotating
models.

For a slowly rotating star, the frequency $\sigma$ of these modes is no longer
zero. Instead, conservation of circulation implies in the Newtonian
limit (Papaloizou and Pringle 1978) the relation
\begin{equation}
\sigma +m\Omega = \frac{2m\Omega}{l(l+1)},
\label{sigma}\end{equation}
where the perturbation's time dependence is $e^{i \sigma t}$.

Eq.\ (\ref{sigma}) is a consequence of conservation of circulation, the curl of
Eq.\ (\ref{euler}):
\begin{equation}
(\partial_t+im\Omega )\epsilon^{abc}\nabla_b\Delta v_c=0.
\end{equation}
(To obtain this equation, note that in Eq. (\ref{euler}) $\delta$ can be 
replaced by $\Delta$, because the equation is satisfied by the equilibrium 
configuration.) 
For $\xi^a$ of the form (\ref{xi}), we obtain
\begin{eqnarray}
\epsilon^{abc}\nabla_b\Delta v_c\nabla_ar &=&
(\partial_t+\Omega\partial_\phi
)\epsilon^{r\theta\phi}(\partial_\theta\xi_\phi -\partial_\phi\xi_\theta
)-2\Omega\sin\theta \xi^\theta\nonumber\\
&=& i [(\sigma +m\Omega) l (l +1)-2m\Omega ] \frac{\zeta}{r^2} Y_\l^m .
\label{canon}\end{eqnarray}
Eq.\ (\ref{canon}) then has as a solution
\begin{equation}
\sigma =-m\Omega \left[ 1-\frac{2}{l(l+1)}\right] .
\label{sigma2}\end{equation}
The frequency $\sigma$ is negative when $l\geq 2$ while the frequency $\sigma +m\Omega$ relative to a comoving observer is positive (Eq. \ref{sigma}).
Because modes that travel backwards relative to the star, are pulled
forwards relative to an inertial frame, they are formally unstable for
arbitrarily small values of the star's rotation.

	Stability is governed by the sign of the canonical energy $E_c$, when
expressed in terms of canonical displacements $\xi^a$ (Friedman and Schutz
 1978).  That is, a model is unstable to perturbations with angular
dependence $e^{im\phi}$ when there is initial data with this angular
behavior for which $E_c(\xi)<0$.  The canonical energy has the form
\begin{eqnarray}
E_c &=& \frac{1}{2} \int [\rho |\dot\xi |^2 - \rho |v\cdot\nabla\xi |^2 +
\gamma p |\nabla\cdot\xi |^2\nonumber\\
& & \mbox{} + (\bar\xi\cdot\nabla p \nabla\cdot\xi + \xi\cdot\nabla p
\nabla\cdot \bar\xi\nonumber\\
& & \mbox{} + \xi^a\xi^b(\nabla_a\nabla_bp+\rho\nabla_a\nabla_b\nu
)-\frac{1}{4\pi} |\nabla\delta\nu |^2]dV .
\label{ec1}\end{eqnarray}
We evaluate it to order $\Omega^2$.

	For a slowly rotating star,
\begin{equation}
\rho =\rho_0 + O(\Omega^2), \hspace{2mm}  \rho =p_0+O(\Omega^2),
\hspace{2mm} \nu =\nu_0+O(\Omega^2),
\end{equation}
with $\rho_0$, $p_0$ and $\nu_0$ values for the corresponding spherical
star with the same mass.  Then for $\xi^a$ of the form (\ref{xi}), we have
\begin{equation}
\xi\cdot\nabla\rho =O(\Omega^2), \hspace{2mm} \xi\cdot\nabla p=O(\Omega^2),
\hspace{2mm} \xi\cdot\nabla\nu = O(\Omega^2) ,
\label{xidot}\end{equation}
and
\begin{equation}
E_c = \frac{1}{2} \int dV\ [\rho |\dot\xi |^2 - \rho |v\cdot\nabla\xi |^2 +
\xi^a\xi^b(\nabla_a\nabla_bp + \rho\nabla_a\nabla_a\nu )] + O(\Omega^4).
\label{ec2}\end{equation}

     In the final term,
\begin{eqnarray}
\nabla_a\nabla_bp + \rho\nabla_a\nabla_b\nu &=& \rho\nabla_a
\left(\frac{1}{\rho} \nabla_bp+\nabla_b\nu\right) + \frac{1}{\rho}
\nabla_a\rho \nabla_bp\nonumber\\
&=& \frac{1}{2} \rho\nabla_a\nabla_b v^2 + \frac{1}{\rho} \nabla_a\rho
\nabla_bp,
\end{eqnarray}
where the equation of hydrostatic equilibrium, (\ref{equil}), is used to obtain
the last equality.  Then
\begin{equation}
\bar\xi^a\xi^b(\nabla_a\nabla_bp+\rho\nabla_a\nabla_b\nu
)=\frac{1}{2}\rho\Omega^2\bar\xi^a\xi^b\nabla_a\nabla_b(\varpi^2),
\end{equation}
with $\varpi =r\sin\theta$ the distance from the symmetry axis.

After a straightforward computation, we find
\begin{equation}
E_c =\frac{1}{2} \int dV\, \rho\left[ |\dot\xi
|^2-m^2\Omega^2|\xi |^2 + 4m\Omega^2r^2\sin\theta |\cos\theta
\xi^\theta \xi^\phi|\right] .
\label{ec3}\end{equation}

	An initial data set $(\xi ,\dot\xi )$ is canonical if it satisfies
\begin{equation}
q^a\equiv \epsilon^{abc}\nabla_b\Delta_\xi v_c=0 .
\end{equation}
Writing $v^a=\Omega\phi^a$ for the equilibrium star and $z^a$ for the unit
vector $\nabla^az$, we obtain
\begin{equation}
q^a=(\partial_t+\Omega\mbox{\pounds}_\phi)\epsilon^{abc}\nabla_b\xi_c
-2\Omega\mbox{\pounds}_z\xi^a
\end{equation}
In a cylindrical chart $\varpi , z, \phi$, the components of $q^a$ are
\begin{equation}
q^i = (\partial_t+\Omega\partial_\phi ) \frac1\varpi
(\partial_j\xi_k-\partial_k\xi_j)-2\Omega\partial_z\xi^i, \hspace{2mm}
i,j,k {\mbox{ cyclic}} .
\end{equation}

	To show the existence of canonical initial data that makes $E_c$ negative
for each $m$ we specialize to the case $l=m$. A displacement $\xi^a$ given by eq.\ (\ref{xi}) with $l=m$ will be canonical
if for suitable $\sigma$,
\begin{equation}
\zeta (r) \propto r^{m+1}, \hspace{2mm} \partial_t\zeta = i\sigma\zeta .
\label{zeta}\end{equation}
We can choose
\begin{eqnarray}
\xi^\varpi &=& i\, \varpi^{m-1}z\, e^{im\phi}\label{xic1}\\
\xi^z &=& -i\, \varpi^m\, e^{im\phi}\label{xic2}\\
\xi^\phi &=& -\, \varpi^{m-2}z\, e^{im\phi} .
\label{xic3}\end{eqnarray}
We then have
\begin{eqnarray}
q^\varpi &=& i[(\sigma +m\Omega )(m+1)-2m\Omega ]\varpi^{m-1}e^{im\phi}\\
q^z &=& 0\\
q^\phi &=& -[(\sigma + m\Omega )(m+1)-2m\Omega ]\varpi^{m-2}e^{im\phi} ,
\end{eqnarray}
and $q^a=0$, when
\begin{equation}
\sigma =-m\Omega \left[ 1-\frac{2}{m(m+1)}\right] .
\label{sigma3}\end{equation}
The agreement with Eq.\ (\ref{sigma2}) reflects the fact that the displacement 
vector of a normal mode is canonical.  

Armed with $\xi^a$, we can verify that the canonical energy is negative. From (\ref{xic1}-\ref{xic3}) we have,
\begin{equation}
|\xi |^2 =  \frac{\zeta^2}{r^2} \sin^{2(m-1)} \theta\ (1+\cos^2\theta ) ,
\end{equation}
and
\begin{equation}
r^2\sin\theta\cos\theta\ |\xi^\theta\xi^\phi | =  \frac{\zeta^2}{r^2}
\sin^{2(m-1)}\theta \cos^2\theta .
\end{equation}
Then, explicitly performing the $\theta$ integration, we find that 
the canonical energy of Eq. (\ref{ec3}) is given by  
\be 
E_c = - \frac{1}{2} \int dV\, \rho |\xi|^2\,\left[ (m\Omega-\sigma)(m\Omega+\sigma) -\frac {2m}{2m+1}\Omega^2\right],
\label{ec4}\end{equation}
implying 
\be 
E_c<0,
\ee
for $\sigma$ satisfying Eq. (\ref{sigma}).

For a normal mode, the canonical condition is equivalent to the curl of 
the perturbed Euler equations.  In fact, the displacement $\xi^a$ given by Eqs. (\ref{xic1}-\ref{xic3}) (or, equivalently, by Eqs. (\ref{xi}) and (\ref{zeta})) satisfies the full Euler equations and is the correct form of an $l=m$ axial mode up to terms of
order $\Omega^3$.

\section{Relativistic stars}
\label{s:gr}

The spacetime of a uniformly rotating relativistic star is described, to lowest order in angular velocity, by 
a metric 
\begin{equation}
ds^2=-e^{2\nu}dt^2+e^{2\psi}(d\phi - \omega dt)^2 +e^{2\lambda} dr^2
+e^{2\mu}d\theta^2,
\label{metric}
\ee
with commuting Killing vectors $t^\alpha$ and $\phi^\alpha$.  The fluid has 
four-velocity
\be
u^\alpha = u^t(t^\alpha + \Omega \phi^\alpha),
\ee
and its energy density $\epsilon$, pressure $p$, and specific entropy $s$, 
satisfy an equation of state,
\be
\epsilon = \epsilon(p,s)
\ee
(neutron stars are effectively isentropic, with no dependence on $s$). 
The field equations
\be
G_{\alpha\beta}=8\pi T_{\alpha\beta}
\ee
imply the equation of hydrostatic equilibrium,
\be
\nabla_\alpha(\tilde h+\log u^t) = 0,
\end{equation}
where
\be
\tilde h = \int \frac{dp}{\epsilon+p}.
\ee

As in the Newtonian limit, an axial perturbation of a spherical relativistic
star involves a time-independent change in the fluid's 4-velocity, and 
no change in the star's pressure or density.  (The metric, however, does 
change, as will be discussed below in the treatment of initial data 
describing a perturbed rotating star.)  The change in the 4-velocity has 
the form 
\begin{equation}
\delta u^\alpha = \zeta(r)\ \epsilon^{\alpha\beta\gamma\delta}
                   \nabla_\beta t\nabla_\gamma r\nabla_\delta Y^l_m,
\end{equation}     
where 
\be
\epsilon^{\alpha\beta} := \epsilon^{\alpha\beta\gamma\delta} u_\gamma
r_\delta ,
\ee
with
\begin{equation}
r_\alpha = e^\lambda\nabla_\alpha r.
\end{equation}

For a rotating star, the frequency $\sigma$ of these modes is
no longer zero.  To show that axial perturbations with arbitrary 
values of $m$ are unstable for slowly rotating perfect-fluid models,
we shall again consider perturbations corresponding
to a Lagrangian displacement with initial data of the form 
\begin{equation}
\xi^\alpha = \zeta (r)\epsilon^{\alpha\beta}\nabla_\beta Y_m^l
\label{xir}\end{equation}
\be
\mbox{\pounds}_{\bf t} \xi^\alpha = i\sigma(r) \xi^\alpha.
\label{xidotr}
\ee
The dependence of $\sigma$ on $r$ arises because it is easier to work
with initial data that is not data for a pure outgoing mode.   

Initial data for the perturbed star and geometry is a set
$(\xi^\alpha, \mbox{\pounds}_{\bf t} \xi^\alpha, h_{\alpha\beta},
\mbox{\pounds}_{\bf t}h_{\alpha\beta})$, where $h_{\alpha\beta}=\delta
g_{\alpha\beta}$, satisfying the initial value equations,
\be
\delta(G^{\alpha\beta} -8\pi T^{\alpha\beta})\nabla_\beta t = 0,
\ee
and the equations governing a canonical displacement.
Given such a set, we have (see Friedman and Ipser 1992 for a review of 
the perturbation formalism used here)
\be
\Delta g_{\alpha\beta} = h_{\alpha\beta} + 2\nabla_{(\alpha}\xi_{\beta )},
\end{equation}
\begin{equation}
\Delta u^\alpha = \frac{1}{2} u^\alpha u^\beta u^\gamma\Delta g_{\beta\gamma},
\end{equation}
and 
\begin{equation}
\frac{\Delta p}{\gamma p} = \frac{\Delta\epsilon}{\epsilon +p} =
\frac{\Delta n}{n} =-\frac{1}{2} q^{\alpha\beta} \Delta g_{\alpha\beta}.
\end{equation}

A spherical star has metric
\be 
ds^2=-e^{2\nu}dt^2+e^{2\lambda} dr^2 +r^2(d\theta^2+\sin^2 \theta d\phi^2)
\ee
and four-velocity
\be
u^\alpha = e^{-\nu} t^\alpha.
\ee
Axial perturbations of the metric can have nonzero vector contributions 
\be
h_{\alpha \beta} t^\beta, h_{\alpha \beta} r^\beta 
\ee
proportional to the vector
\be
\epsilon_\alpha^\beta\nabla_\beta Y_l^m,
\ee
and a nonzero tensor contribution proportional to
\be 
\epsilon_{(\alpha}^\gamma \nabla_{\beta)}\nabla_\gamma Y_l^m.
\ee

The form of $\xi^\alpha$ implies 
\begin{equation}
\Delta u^\alpha =0,
\end{equation}
\begin{equation}
\delta u^\alpha =-e^{-\nu}\mbox{\pounds}_{\bf t} \xi^\alpha,
\end{equation}
and
\begin{equation}
\delta\epsilon = \delta p = 0.
\end{equation}

A slowly rotating star has metric for which $\omega$ is $O(\Omega)$ and
the remaining departure from spherical is $O(\Omega^2)$:
\be
ds^2=-e^{2\nu}(1+2h)dt^2+r^2(1+2k)\sin^2 \theta( d\phi-\omega dt)^2+e^{2\lambda} (1+2\ell) dr^2 +r^2(1+2k) d\theta^2;
\ee
we will need only the fact that $h,k,\ell$ are $O(\Omega^2)$, not their
explicit values (Hartle 1967).  
For a displacement of the form (\ref{xir}), we have
\begin{equation}
\Delta u^\alpha = O(\Omega^2)
\end{equation}
\begin{equation}
\delta\epsilon = O(\Omega^2), \hspace{3mm} \delta p = O(\Omega^2)
\end{equation}

The condition governing a canonical displacement is again related to 
conservation of circulation, which has the relativistic form (Friedman 1978)
\be
\mbox{\pounds}_{\bf u} \omega_{\alpha\beta} = 0,  \label{circ}
\ee
where $\omega_{\alpha\beta}$ is the relativistic vorticity,
\be
\omega_{\alpha\beta} = 2 \nabla_{[\alpha}\left( \frac{\epsilon +p}{n}
u_{\beta ]}\right) 
\label{vort} \ee
Then canonical initial data satisfies  
\begin{equation}
\Delta\omega_{\alpha\beta} =0
\label{canonr}\end{equation}
Now
\begin{equation}
\Delta\omega_{\alpha\beta} = 2 \nabla_{[\alpha}  
              \left(\frac{\epsilon +p}{n} \Delta u_{\beta ]}\right) + 	
		O(\Omega^2),
\label{domega}\end{equation} 
with 
\begin{eqnarray}
\Delta u_\alpha &=& \Delta (g_{\alpha\beta}u^\beta )=\Delta g_{\alpha\beta}
u^\beta + O(\Omega^2)\nonumber\\
&=& 2\nabla_{(\alpha}\xi_{\beta )} u^\beta + O(\Omega^2) .
\end{eqnarray}

As in the Newtonian case, we will use data with $l = m$, writing $\xi^\alpha$ 
in the form
\begin{equation}
\xi^\theta = i\sin^{m-1}\theta\, e^{im\phi} \frac{\zeta(r,t)}{r^2}
\end{equation}
\begin{equation}
\xi^\phi =-\sin^{m-2}\theta\, e^{im\phi} \frac{\zeta(r,t)}{r^2} \cos\theta
\end{equation}
\begin{equation}
\mbox{\pounds}_{\bf t}\xi =i\sigma\xi .
\end{equation}
From Eq. (\ref{domega}), we obtain 
\begin{eqnarray}
\Delta\omega_{\theta\phi} &=& 2 \partial_{[\theta}\left( \frac{\epsilon
+p}{n} \Delta u_{\phi ]}\right) = 0 :\nonumber\\
& & u^t \sin^{m+1}\theta\, e^{im\phi} \zeta [-im(m+1)(\sigma +m\Omega
)+2im(\Omega-\omega )] = 0,
\end{eqnarray}
satisfied when 
\begin{equation}
\sigma + m\Omega = \frac{2m(\Omega -\omega )}{m(m+1)}.
\label{sigmar}\end{equation}
Thus, as noted earlier, we have not supplied data for a normal mode, 
but data for which $\sigma$ is a function of $r$, with
\begin{equation}
\partial_r\sigma =-\frac{2}{m+1} \partial_r\omega.
\end{equation}

The remaining components of the canonical condition (\ref{canonr}) are
\begin{eqnarray}
& & \Delta \omega_{r\phi} = 0:\nonumber\\
& & \mbox{} \partial_r \log \left( \frac{\epsilon +p}{n} \right)
\Delta u_\phi + \partial_r \left(\Delta u_\phi\right) -\partial_\phi\left(  \Delta u_r\right) =0,
\end{eqnarray}
and 
\begin{eqnarray}
& & \Delta \omega_{r\theta} = 0:\nonumber\\
& & \mbox{} \partial_r \log \left( \frac{\epsilon +p}{n} \right)
\Delta u_\theta + \partial_r \left(\Delta u_\theta\right) -\partial_\theta\left(  \Delta u_r\right) =0.
\end{eqnarray}
Using the equation of hydrostatic equilibrium in the form 
\[
	\partial_r\left( \frac{ \epsilon+p}{nu^t} \right) = 0 ,
\]
one obtains from the remaining components of the canonical condition   a
 single equation 
for $\zeta$,
\begin{equation}
\partial_r\zeta \frac{2(\Omega -\omega )}{m+1}
 + \zeta \left[ - \frac{2}{m+1} \partial_r\omega -  \frac{1}{r^2}
\partial_r (r^2(\Omega -\omega )) - 2\partial_r\nu \left( \frac{2(\Omega
-\omega )}{m+1} -(\Omega-\omega)\right)\right] = 0,
\end{equation}
with solution,
\begin{equation}
\zeta = (\Omega -\omega )^{-1+(m+1)/2} r^{m+1} \exp\left[ -(m-1) \nu \right].
\end{equation}

We have obtained a canonical displacement, but must still satisfy the 
initial value equations.  We will do so by choosing a metric perturbation 
for which $h_{\alpha\beta} = 0$, $\mbox{\pounds}_{\bf t} h_{\alpha\beta} \neq 0$.
The initial value equations can be obtained from the action
for the perturbation equations, which we will use in any event to
write down the canonical energy: 
\be
I = \int d^4x \sqrt{-g} {\cal L}(\bar\xi ,\bar h;\xi ,h)
\ee
where
\ba {\cal L} (\bar\xi ,\bar h;\xi ,h) &=& U^{\alpha\beta\gamma\delta}
\nabla_\alpha\bar\xi_\beta\nabla_\gamma\xi_\delta +
V^{\alpha\beta\gamma\delta}(\bar
h_{\alpha\beta}\nabla_\gamma\xi_\delta+h_{\alpha\beta}\nabla_\gamma\bar\xi_\delta)\cr
&&\quad - {1\over{32\pi}} \epsilon^{\alpha\gamma\epsilon\eta}\epsilon^{\beta\delta\zeta}{_\eta}\nabla_\gamma\bar h_{\alpha\beta}\nabla_\delta h_{\epsilon\zeta}\cr
&&\quad
-T^{\alpha\beta}R_{\alpha\gamma\beta\delta}\bar\xi^\gamma\xi^\delta +
\left( {1\over 2} W^{\alpha\beta\gamma\delta}-{1\over{16\pi}}
G^{\alpha\beta\gamma\delta}\right) \bar h_{\alpha\beta}h_{\gamma\delta}\cr
&&\quad - {1\over 2} \nabla_\gamma T^{\alpha\beta}(\bar h_{\alpha\beta}\xi^\gamma+h_{\alpha\beta}\bar\xi^\gamma) ,\label{lagr}\ea
with
\ba
G^{\alpha\beta\gamma\delta} = {1\over 2} R^{\alpha(\gamma\delta)\beta} &+& {1\over 4}
(2R^{\alpha\beta}g^{\gamma\delta}+2R^{\gamma\delta}g^{\alpha\beta}-3R^{\alpha(\gamma}g^{\delta)\beta}-3R^{\beta(\gamma}g^{\delta)\alpha})\cr
&+& {1\over 4} R(g^{\alpha\gamma}g^{\beta\delta}+g^{\alpha\delta}g^{\beta\gamma}-g^{\alpha\beta}g^{\gamma\delta}) , 
\label{gabcd}\ea
\be
U^{\alpha\beta\gamma\delta} = (\epsilon +p)u^\alpha u^\gamma q^{\beta\delta}+p(g^{\alpha\beta}g^{\gamma\delta}-g^{\alpha\delta}g^{\beta\gamma})-\gamma p\;
q^{\alpha\beta}q^{\gamma\delta} , 
\label{uabcd}\ee
and
\be
2V^{\alpha\beta\gamma\delta} = (\epsilon +p)(u^\alpha u^\gamma q^{\beta\delta}+u^\beta u^\gamma q^{\alpha\delta}-u^\alpha u^\beta q^{\gamma\delta}) - {1\over 2} \gamma
p\; q^{a\beta}q^{\gamma\delta} . 
\label{vabcd}\ee

The perturbed initial value equations are given by 
\be
0 = \frac{\delta I}{\delta h_{\alpha\beta}} \nabla_\beta t. 
\ee
When $h_{\alpha\beta}$ vanishes on the intial data surface, the Hamiltonian constraint vanishes. We choose $\dot h_{\alpha\beta}$ 
to have no vector contribution, writing
\begin{equation}
\dot h_{\alpha\beta} = h(r)\epsilon_{(\alpha}{}^\gamma \nabla_{\beta
)}\nabla_\gamma Y_l^m.
\end{equation}
The momentum constraint then has, to $O(\Omega)$, the form 
\begin{equation}
\frac12(\epsilon +p) u^t\mbox{\pounds}_{\bf u}\xi^\alpha -\frac{1}{32\pi}
\epsilon^{t\alpha\gamma\zeta} \epsilon^{t\beta\delta}{}_\zeta \nabla_\beta
\dot h_{\gamma\delta} = 0,
\end{equation}
or
\begin{equation}
i(\sigma +m\Omega )(\epsilon +p)(u^t)^2 \frac{\zeta}{r^2} + \frac{1}{32\pi}
(m+1)(m+2) \frac{e^{-2\nu}}{r^4} h=0.
\end{equation}
Thus the initial value equations are satisfied by choosing 
\begin{equation}
h(r) =-32\pi i \frac{\sigma +m\Omega}{(l-1)(l+2)} r^2 (\epsilon +p)\zeta.
\end{equation}

We can now compute the canonical energy, whose relativistic form is  
\be
E_c = \int_S dS_\epsilon
[U^{\epsilon\beta\gamma\delta}\dot\xi_\beta\nabla_\gamma\xi_\delta + V^{\gamma\delta\epsilon\beta}h_{\gamma\delta}\xi_\beta 
-{1\over 32 \pi }\epsilon^{\epsilon\gamma\rho\zeta} \epsilon^{\beta\delta\eta}{_{\zeta}}\dot h_{\gamma\delta}\nabla_\beta h_{\rho\eta}
- t^\epsilon{\cal L}] .
\label{ecr1}\ee
For a displacement having the axial form (\ref{xir}), we obtain the simpler expression  
\begin{eqnarray}
E_c &=& \int dV e^\nu [U^{t\beta\gamma\delta} 
         \mbox{\pounds}_{\bf t}\bar\xi_\beta \nabla_\gamma\xi_\delta -  
     \frac{1}{64\pi} \epsilon^{t\gamma\delta\zeta}
     \epsilon^{t\delta\eta}{_\zeta} \mbox{\pounds}_{\bf t}\bar
     h_{\gamma\delta}\mbox{\pounds}_{\bf t}h_{\epsilon\eta}\nonumber\\
& & \mbox{} -\frac{1}{2} U^{\alpha\beta\gamma\delta}
\nabla_\alpha\bar\xi_\beta\nabla_\gamma\xi_\delta +\frac{1}{2}
T^{\alpha\beta} R_{\alpha\gamma\beta\delta} \bar\xi^\gamma\xi^\delta ]\cr
&\equiv& \int dV[{\cal E}_{f1} + {\cal E}_{f2}+{\cal E}_{f3}+
{\cal E}_g ],
\end{eqnarray}
where the terms ${\cal E}_{fi}$ indicate contributions from the 
fluid, ${\cal E}_g$ the contribution from the perturbed metric.
(This split into fluid and metric is gauge-dependent).  Evaluating
each term, we obtain 
\begin{equation}
{\cal E}_{f1} = U^{t\beta\gamma\delta} \mbox{\pounds}_{\bf t}\bar\xi_\beta\nabla_\gamma\xi_\delta
=-i\sigma (\epsilon +p)u^t \bar\xi_\beta u^\gamma\nabla_\gamma\xi^\beta
+O(\Omega^4),
\end{equation}
\begin{eqnarray}
{\cal E}_{f2}=-\frac{1}{2} U^{\alpha\beta\gamma\delta} \nabla_{\alpha} \bar\xi_\beta
\nabla_\gamma\xi_\delta = &-& \frac{1}{2} (\epsilon +p) |u\cdot\nabla\xi
|^2 + \frac{1}{2} p\nabla_\alpha\bar\xi_\beta
\nabla^\beta\xi^\alpha\nonumber\\
& & \mbox{} + O(\Omega^4) .
\end{eqnarray}
\begin{eqnarray}
{\cal E}_{f3} = \frac{1}{2}
T^{\alpha\beta}R_{\alpha\gamma\beta\delta}\bar\xi^\gamma\xi^\delta =
&-&\frac{1}{2} (\epsilon +p) u^\beta\nabla_\beta\bar\xi^\gamma
u^\delta\nabla_\gamma\xi_\delta -\frac{1}{2} p
\nabla_\alpha\bar\xi_\beta\nabla^\beta\xi^\alpha\nonumber\\
&+& \frac{1}{2} (\epsilon
+p)(u^\beta\nabla_\alpha\xi_\beta\bar\xi^\gamma\nabla_\gamma u^\alpha
+u^\alpha\nabla_\alpha\xi_\beta\bar\xi^\gamma\nabla_\gamma u^\beta
)\nonumber\\
&+& e^{-\nu} D_\alpha (e^\nu T^{\gamma[\alpha}\bar\xi^{\beta]}
\nabla_\beta\xi_\gamma ).
\end{eqnarray}
In this final equality, $D_\alpha$ is the derivative operator with
respect to the 3-metric on the $t=$constant initial value surface.
Then
\begin{eqnarray}
{\cal E}_f&\equiv &{\cal E}_{f1} + {\cal E}_{f2}+{\cal E}_{f3} \cr 
&=& \frac{1}{2} (\epsilon +p) u^\beta\nabla_{(\alpha}\xi_{\beta )}
\mbox{\pounds}_{\bf u} \bar\xi^\alpha - i\sigma (\epsilon +p)u^t \bar\xi_\beta
u^\gamma\nabla_\gamma\xi^\beta\nonumber\\
& &\mbox{} + e^{-\nu} D_\alpha (e^\nu
T^{\gamma[\alpha}\bar\xi^{\beta]} \nabla_\beta\xi_\gamma )\nonumber\\
& = & - \frac{i}{2} (\sigma -m\Omega )(\epsilon +p) u^t \bar\xi_\beta
u^\gamma\nabla_\gamma\xi^\beta + \frac{i}{2} (\sigma + m\Omega )(\epsilon +p) u^t \bar\xi^\alpha\xi^\beta\nabla_\alpha u_\beta\nonumber \\
&& + e^{-\nu} D_\alpha A^\alpha,
\end{eqnarray}
where
\begin{equation}
A^\alpha =  e^\nu [T^{\gamma[\alpha}\bar\xi^{\beta]} \nabla_\beta\xi_\gamma + \frac{i}{2}(\sigma +m\Omega )(\epsilon +p)u^t \bar\xi^\alpha u^\beta\xi_\beta ].
\end{equation}

The remaining term is 
\begin{eqnarray}
{\cal E}_g&= & -\frac{1}{64\pi}\epsilon^{t\gamma\delta\zeta}
	\epsilon^{t\epsilon\eta}{_\zeta}
      \, \mbox{\pounds}_{\bf t}\bar
h_{\gamma\epsilon}\mbox{\pounds}_{\bf t}h_{\delta\eta} = 
\frac{1}{64\pi} e^{-2\nu}\,
e^{\alpha\beta}\, e^{\gamma\delta} \mbox{\pounds}_{\bf t}\bar
h_{\alpha\gamma}\mbox{\pounds}_{\bf t} h_{\beta\delta},
\end{eqnarray}
where $e_{\alpha\beta}$ is the metric on a $2$-sphere of radius $r$,
\be
\mbox{} e_{\alpha\beta} = r^2(\nabla_\alpha\theta\,
\nabla_\beta\theta + \sin^2\theta\, \nabla_\alpha\phi\, \nabla_\beta\phi ).
\ee
Thus 
\be
E_c = \int dV({\cal E}_f+{\cal E}_g), 
\ee
where the terms have opposite signs:
\ba
{\cal E}_f &=& 
-\frac{1}{2} (m\Omega - \sigma )(m\Omega + \sigma)e^{-2\nu}(\epsilon +p)|\xi|^2
                                                                  \nonumber \\ 
 && + 2m\Omega(\Omega-\omega)e^{-2\nu}(\epsilon+p)
                              r^2\sin\theta|\cos\theta \xi^\theta\xi^\phi| <0
	\label{efluid}\\
{\cal E}_g &=& \frac{1}{64\pi} e^{-2\nu} e^{\alpha\beta} e^{\gamma\delta}
{\dot{\bar h}}_{\beta\delta} \dot h_{\alpha\gamma}>0.
\ea
The fluid contribution ${\cal E}_f$ has the same form as the canonical energy 
(\ref{ec3}) for Newtonian modes, from which the sign of (\ref{efluid}) follows.

The sign of $E_c$ is negative because the fluid contribution is dominant.  
Writing 
\ba
{\cal E}_f &=& -\frac{2}{m+1}(\Omega -\omega) \left[\left(m\Omega-\frac{\Omega-\omega}{m+1}\right) -\cos^2\theta \left(m^2\Omega+\frac{\Omega-\omega}{m+1}\right)\right]\times\nonumber\\ 
&& e^{-2\nu} (\epsilon +p) \frac{\zeta^2}{r^2} \sin^{2m-2}\theta 
                \\
{\cal E}_g &=& 32\pi \left[ \frac{m}{(m+1)(m+2)}\right]^2 (\Omega -\omega )^2 	
							\nonumber\\
     & & \mbox{}  e^{-2\nu}(\epsilon +p)^2\zeta^2
         \sin^{2m}\theta \left( 1+\frac{8\cos^2\theta}{\sin^4\theta} \right),
\ea
we have
\be
\frac{\int d\Omega {\cal E}_g}{|\int d\Omega {\cal E}_f|} =
\frac{16}{[(m-1)(m+2)]^2}\ 2\pi (\epsilon +p)r^2 < 2\pi (\epsilon +p)r^2.
\ee
The maximum value of $2\pi (\epsilon +p)r^2$,  for any equation of state, occurs for the maximum mass star. We directly  calculate  $2\pi (\epsilon +p)r^2$ 
for the maximum mass star using equations of state spanning a wide range of 
stiffness.  For equations of state G (excessively soft), C and L (excessively stiff) from the Arnett and Bowers (1977) catalogue we find 
\be
2\pi (\epsilon +p)r^2 < 0.4, 
\ee
for all radius, implying 
\be E_c <0.
\ee

\section{Discussion}
\label{s:time}
 Although a mode with negative canonical energy is unstable, the
 growth times of axial modes appear to be proportional to a high power
 of $\Omega$ for slowly rotating stars. Because the nonaxisymmetric
 instability of rotating perfect-fluid models is driven by
 gravitational radiation, its growth time for modes with a given $m$
 is ordinarily shortest for modes that correspond to the smallest
 value of $l$ for a given $m$, namely $l=|m|$.  For axial
 perturbations of slowly rotating stars, however, modes corresponding
 to a given $l$ and $m$ for the spherical star give rise to density
 perturbations associated with $Y_{l\pm 1}^m$. For nonisentropic
 stars, axial modes with $l=|m|+1$ give rise to $Y^m_l$ density
 perturbations, and as a result, these modes have the shortest growth
 time.  However, if the star is isentropic, it can be shown (Provost
 et. al. (1981)) that the only axial normal modes of a Newtonian star
 have $l=|m|$, which contribute a $m+1$ multipole to the density perturbation. 
As a result, the gravitational radiation from mass multipoles is dominated
by radiation from current multipoles (Lindblom, Owen \& Morsink 1998).

We provide here a rough estimate of the time scale for the axial
instability of a mode with angular dependence $\exp(im\phi)$, for a
slowly rotating, nearly Newtonian model.  A mode characterized by a
displacement vector $\xi$ has energy of order
\begin{equation}
E\sim -M\Omega^2 \xi^2 .
\end{equation}
The mode is driven by gravitational radiation, with power
dominated by the lowest order current multipole
 of order
\begin{equation}
\frac{dE}{dt} \sim -\left( \frac{d^{l+1}}{dt^{l+1}} J^m_l\right)^2
\hspace{2mm} \sim -\Omega^{2l+2}| J^m_l|^2
\end{equation}
where $J^m_l$ is the lowest current multipole  perturbation 
of the star.  We have
\begin{equation}
J^m_l\sim\int \rho \;  \delta v^a \epsilon_a{^{bc}} \nabla_b r \nabla_c
	Y^m_l r^l dV .
\end{equation}
With our choice of $\xi$, $\delta v^a$ is $O(\Omega),$\footnote{It would
be more natural from a physical standpoint to take $\delta v^a = 
O(1)$, $\xi = O(\Omega^{-1})$ because as $\Omega \rightarrow 0$, the 
perturbation goes over to a nonvanishing, time-independent perturbation
 of the spherical star.} 
\begin{equation}
J^m_l \sim\Omega \xi R^{l+2}
\end{equation}
Finally, the growth time is of order
\begin{eqnarray}
\tau &=& E/\frac{dE}{dt} =
\frac{M\Omega^2\xi^2}{\Omega^{2m+2}|J^m_m|^2}\nonumber\\
&\sim& \frac{M}{(R\Omega )^{2m+2}} .
\end{eqnarray}
The precise calculation is given by Lindblom et. al. (1998).
The timescale appears to have the same
$\Omega$ dependence for relativistic stars 
(Friedman, Lockitch \& Morsink 1998).

The imaginary part of the frequency that Andersson (1997) computes is
$O(\Omega^2)$, and our timescale estimate suggests that its magnitude is
an artifact of an $O(\Omega^2)$ approximation method.  But the sign of the
imaginary part that his method obtains may correctly diagnose instability
of a mode in the slow-rotation approximation.

The discussion so far has ignored viscosity, and, as in the case of
polar modes, a long growth time for slow rotation implies that
viscosity will enforce stability except for hot, rapidly rotating
neutron stars (see Lindblom and Mendell 1995 and references therein).
In particular, because the time-independent axial modes of a spherical
star have nonvanishing shear, viscous damping must damp the instability
for slow rotation.  

For rapidly rotating stars, however, instability points have been
computed numerically only for polar modes, initially for Newtonian
models (Managan 1985, Imamura et al. 1985) and more recently for
relativistic models in a Cowling approximation by Yoshida and Eriguchi
(1997) and in the full theory by Stergioulas and Friedman (Stergioulas
1996, Stergioulas and Friedman 1997). Estimates of instability points
when realistic values viscosity is included have similarly been
restricted to polar perturbations.  Extending these estimates to axial
perturbations should then be straightforward.  But for both axial and
polar modes, a precise computation of instability points when viscosity
is included is a more difficult task, apparently requiring the
construction of outgoing modes and determination of their complex
frequencies.
   
\acknowledgments
We are grateful to Nils Andersson for acquainting us with his results
and for several helpful discussions.  This work was supported in part by 
NSF Grant No. PHY95-07740 and by NSERC of Canada.



\begin{references}

\reference{} Andersson, N. 1997 ``A new class of unstable modes of 
rotating relativistic stars,'' ApJ, to appear (gr-qc/9706075)  

\reference{Arnett} Arnett, W.D., \& Bowers, R. L. 1977, ApJ Sup., { 33}, 415

\reference{Chandra70} Chandrasekhar, S. 1970, Phys. 
   Rev. Lett., { 24}, 611

\reference{Friedman78} Friedman, J. L., 1978, Commun. Math. Phys., { 62}, 247

\reference{FriedmanIpser92} Friedman, J. L., \& Ipser, 
   J. R. 1992, Phil. Trans. R. Soc. Lond., { A340}, 391

\reference{} Friedman, J.L., Lockitch, K.H., \& Morsink, S.M. 1998,
	paper in preparation

\reference{F&S75b} Friedman, J. L. \& Schutz, B. F. 
   1975, ApJ, { 200}, 204 

\reference{F&S78a} Friedman, J. L  \& Schutz, B. F. 
   1978, ApJ, { 221}, 937 

\reference{F&S78b} Friedman, J. L. \& Schutz, B. F. 
   1978, ApJ, { 222}, 281

\reference{IFD85} 
   Imamura, J. N., Friedman, J. L. \& Durisen, R. H. 1985,
   ApJ, { 294}, 474

\reference{hartle} Hartle, J.B., 1967, ApJ, 150, 1005 

\reference{} Lindblom, L. \& Mendell, G., 1995, ApJ, 444, 809

\reference{} Lindblom, L., Owen, B.J., \& Morsink, S.M. 1998,
	Phys. Rev. Lett. 80, 4843 (gr-qc/9803053)

\reference{Managan85} Managan, R. A. 1985, ApJ, { 294}, 463.

\reference{Papaloizou} Papaloizou, J. \& Pringle, J. E., 1978,
	Mon. Not. R. astr. Soc., 182, 423. 

\reference{Provost} Provost, J., Berthomieu, G. and Rocca, A., 1981,
	Astron. Astrophys. 94, 126. 

\reference{ReggeWheeler57} Regge, T. \& Wheeler, J. A. 
   1957, Phys. Rev. { 108}, 1063

\reference{Sterg} Stergioulas, N., 1996, {The Structure and Stability of 
	Rotating Relativistic Stars}, Ph. D. Thesis, University of 
	Wisconsin-Milwaukee. 

\reference{StergFriedman} Stergioulas, N. \& Friedman, J. L., 1998,
	ApJ, 492, 301 

\reference{YoshidaEriguchi95} Yoshida, S. \& Eriguchi, Y. 1997, 
	ApJ, 490, 779

\end{references}
\end{document}